# Inlaid ReS$_2$ Quantum Dots in Monolayer MoS$_2$

*Ziqian Wang,[1] Ruichun Luo,[1] Isaac Johnson,[1] Hamzeh Kashani,[1] and Mingwei Chen,[1,2,*]*

[1] Department of Materials Science and Engineering, Johns Hopkins University, Baltimore, MD 21218, USA; [2] WPI Advanced Institute for Materials Research, Tohoku University, Sendai 980-8577, Japan.

*Address correspondence to: mwchen@jhu.edu

ABSTRACT: Two-dimensional (2D) transition metal dichalcogenides (TMDs) are prospective materials for quantum devices owing to their inherent 2D confinements. They also provide a platform to realize even lower-dimensional in-plane electron confinement, *e.g.*, 0D quantum dots, for exotic physical properties. However, fabrication of such laterally confined monolayer (1L) nanostructure in 1L crystals remains challenging. Here we report the realization of 1L ReS$_2$ quantum dots epitaxially inlaid in 1L MoS$_2$ by a two-step chemical vapor deposition method combining with plasma treatment. The lateral lattice mismatch between ReS$_2$ and MoS$_2$ leads to size-dependent crystal structure evolution and in-plane straining of the 1L ReS$_2$ nanodots. Optical spectroscopies reveal the abnormal charge transfer between the 1L ReS$_2$ quantum dots and the MoS$_2$ matrix, most likely resulting from electron trapping in the 1L ReS$_2$ quantum dots. This study may pave the way for realizing in-plane quantum-confined devices in 2D materials for potential applications in quantum information.

KEYWORDS: 2D materials, transition metal dichalcogenide alloy, phase engineering, phase transition, chemical vapor deposition

The emergence of two-dimensional (2D) materials gives rise to a wide variety of possibilities for designing and fabricating low-dimensional quantum devices owing to their inherently confined nature in two-dimension as well as their attractive material properties and intriguing physics.[1] For example, 2D transition metal dichalcogenides (TMDs) have the almost highest light absorption coefficients in all known materials.[2] Their large exciton binding energies, comparable to III-V semiconductors and organic semiconductors,[3] are highly desirable for quantum optoelectronic devices.[4] 2D TMDs are also robust candidates as hosts of valley-spin qubits because of the strong spin-orbit coupling along with the valley degree of freedom.[4–6] Moreover, vertical and lateral heterostructures by the rich combinations of 2D materials further provide high flexibility for the design of quantum confined nanostructures. Especially, lateral heterostructures are expected to realize even lower-dimensional confinement, *e.g.*, 1D quantum wires and 0D quantum dots, on the basis of such naturally out-of-plane confined 2D materials.

Great advances have been made recently on the nanofabrication of 2D materials, such as epitaxy of strain-engineered coherent 1L TMD superlattices, vapor-liquid-solid growth of 1L MoS$_2$ nanoribbons,[7] dislocation-catalyzed or dislocation-driven growth of nanochannels in laterally heterostructured 1L TMDs,[8,9] and embedment of graphene quantum dots in h-BN by Pt nanoparticle-catalyzed conversion or focused helium ion beam nanofabrication.[10,11] Nevertheless, the fabrication of quantum-confined 0D TMD heterostructures with sub-nano-scale size control in lateral dimensions are highly desired for potential applications in quantum information[12] but have not been realized experimentally. Therefore, developing effective nanofabrication methods for such nanosized lateral heterostructures is in urgent need and the knowledge about the structures and properties of the 0D nanostructures is of significance for materials and device designs.

In this study, we demonstrate the successful inlay of ReS$_2$ quantum dots in the MoS$_2$ matrix in monolayer (1L) regime using a two-step low-pressure (LP) chemical vapor deposition (CVD) method assisted with plasma treatment. With type I band alignment relation between pristine 1L ReS$_2$ and 1L MoS$_2$,[13,14] the monolayer MoS$_2$-ReS$_2$ system can potentially produce the 0D confinement in 2D quantum wells. High-angle annular dark-field (HAADF) scanning transmission electron microscopy (STEM) observations show that the epitaxial 1L ReS$_2$ nanodots are achievable with sizes ranging from sub-nanometers to tens of nanometers circumscribed by atomically sharp interfaces. -Atomic structure analyses of the inlaid ReS$_2$ nanodots unveils notable size-dependence in crystal structures and confinement strains which deviate from the continuum mechanical model. Moreover, Raman scattering and photoluminescence reveal the unusual charge transfer between the 1L MoS$_2$ matrix and the 1L ReS$_2$ quantum dots.

RESULTS AND DISCUSSION

Figure 1 schematically illustrates the fabrication process for the inlaid 1L ReS$_2$ nanodots in 1L MoS$_2$ by a two-step CVD. As the first step, 1L MoS$_2$ samples are grown on soda-lime glass substrates by LPCVD at the growth temperature of 750°C using MoO$_3$ and S powders as sources[15,16] (see Methods for details of CVD conditions). The as-grown 1L MoS$_2$ samples, along with the glass substrates, are then exposed to air plasma for 10, 20, and 30 minutes right after the growth (Figure 1a). In this process, a small amount of the Mo and S atoms escape from the 2D lattice under the bombardment of ions, atoms and molecules in the plasma, resulting in Mo- and S-vacancies and the formation of atomic- and nano-scale voids in the 1L MoS$_2$ as the plasma treatment time is prolonged.[17] Such plasma-induced voids with S-zigzag edges can be evidenced by STEM observations, and their sizes and morphologies are shown in Figure S1 (Supporting Information). After the plasma treatment, the defective MoS$_2$ samples are immediately put into a CVD chamber followed by the second CVD step, in which 1L ReS$_2$ is deposited to exactly fill the voids in the 1L MoS$_2$ matrix (Figure 1b). The ReS$_2$ deposition is carried out at the temperature of 700°C using Re$_2$O$_7$ and S powders as precursors while keeping a S-rich environment throughout the growth. A strict 2D lateral growth manner along the edges of the MoS$_2$ voids can be attained by devised control of LPCVD conditions. A sufficiently low Re-precursor concentration in the gas environment is generated to forbid the direct nucleation of ReS$_2$ islands on the surface of 1L MoS$_2$. Instead, the epitaxy of 1L ReS$_2$ is initiated by the formation of Re-S bonds with the unsaturated S at the inner edges of the voids of plasma-treated 1L MoS$_2$, and followed by 1L ReS$_2$ growth until the voids are filled. At the same time, substitution of interface Mo by Re, and relaxation of the interface structure may take place in the ReS$_2$ deposition process. Even though S is possibly removed during the plasma treatment, the S-excessive environment in the second CVD step can repair the S-vacancies and improve the crystallinity of the plasma treated 1L MoS$_2$ matrix.

Different from the stable 2H structure of 1L MoS$_2$, the stable phase of 1L ReS$_2$ in ambient condition has the distorted 1T structure, featured by the periodic striping in one of the in-plane directions.[18–20] Atomic models of 2H MoS$_2$ and distorted 1T ReS$_2$ are shown in Figure S2. The atomic structure of the ReS$_2$ nanodots, embedded in the 1L MoS$_2$ matrix, was characterized using HAADF-STEM as shown in Figure 2. Upper panels of Figure 2a-d are exemplary images of ReS$_2$ nanodots of different sizes, and the corresponding lower panels display the images after deconvolution for sharper view of individual atom columns. Due to the atomic number dependence of the contrast (*Z*-contrast) in HAADF imaging mode, the atoms with the brightest intensities are Re atoms ($Z = 75$), while those with dimmer intensities at the transition metal sites are Mo atoms ($Z = 42$). The overlapping double S atoms ($Z = 16$) show slightly lower intensity than a single Mo atom, verifying that the 1L MoS$_2$ matrix is the semiconducting 2H phase.[21,22] Therefore, the sharp difference in intensities between the two transition metal elements unambiguously distinguishes the ReS$_2$ nanodots and MoS$_2$ matrix as bright and dark regions in the HAADF-STEM images. Each bottom panel of Figure 2a-d displays two intensity profiles along the straight lines between two arrowheads in the raw images (between cyan arrow heads) and the deconvoluted images (between magenta arrowheads), respectively. From the HAADF-STEM images along with their corresponding intensity profiles, it can be explicitly observed that the structure of the 1L $ReS_2$ nanodots evolves with their sizes. Sub-nanometer-sized $ReS_2$ (Figure 2a) accommodates itself to the 2H structure fully coherent with $MoS_2$ matrix as evidenced by the superposing $S_2$ as the feature of the 2H phase in the profiles. A quasi-gap with a slightly large interspacing appears in the $ReS_2$ dot with the diameter merely above 1 nm (Figure 2b) and notes the initiation of the distorted 1T structure of $ReS_2$. Well-defined periodic gaps appear in the $ReS_2$ dot with a size of 2~3 nm (Figure 2c) as the feature of distorted 1T $ReS_2$ lattice. Triangle-shaped $ReS_2$ dots with a single domain form up to lateral size of 5 nm (Figure 2d,h) as the consequence of strong faceting tendency of $ReS_2$-$MoS_2$ interfaces along the S-zigzag edges on the $MoS_2$ side. Furthermore, the interfaces between $ReS_2$ and $MoS_2$ tend to be compact because the characteristic gaps of distorted 1T structure always appear inside the $ReS_2$ nanodots whereas no gaps are observed at $ReS_2$/$MoS_2$ interfaces. In addition, when the size is larger than 6 nm, the $ReS_2$ nanodots always have a multi-domain structure, which will be discussed later.

Based on the observations of thousands of ReS$_2$ nanodots by HAADF-STEM, the ReS$_2$ nanodots are confirmed to be highly coherent with the MoS$_2$ lattice with exceptions less than 0.1%. The example of an exceptional incoherent ReS$_2$ nanodot is shown in Figure S3.

The crystal structure of distorted 1T ReS$_2$ belongs to the triclinic $P\bar{1}$ space group with the lattice parameters of $a(\text{ReS}_2) = 6.576\,\text{Å}$ and $b(\text{ReS}_2) = 6.410\,\text{Å}$ (Re-chain direction) (as well as $|\boldsymbol{a}+\boldsymbol{b}|(\text{ReS}_2) = 6.515\,\text{Å}$) in the basal plane,[19] while the primitive 1L MoS$_2$ in the hexagonal $P\bar{6}m2$ space group is isotropic in the basal plane with the lattice parameter of $a(\text{MoS}_2) = 3.16\,\text{Å}$ (Figure S2).[23] Taking twice of $a(\text{MoS}_2)$ as the reference, the lattice mismatches in $\boldsymbol{a}$, $\boldsymbol{b}$ and $|\boldsymbol{a}+\boldsymbol{b}|$ directions of ReS$_2$ are considered to be 3.8%, 1.4%, and 3.1%, respectively, resulting in different lattice strains for ideally coherent interfaces in the three inequivalent orientations. To study the strain distributions in the inlaid ReS$_2$ nanodots, geometric phase analysis (GPA) is applied to HAADF-STEM images (Figure 3). Figure 3a shows a deconvoluted HAADF-STEM image of a region containing multiple ReS$_2$ nanodots with sizes from ~2 nm to ~17 nm. The three largest nanodots are numbered and the zoom-in image of nanodot #2, indicated by the yellow dashed square, is displayed in Figure 3b. Masks with radii of $|\boldsymbol{g}|/8$, centered at the mutual Bragg spots ($\boldsymbol{g}$) for MoS$_2$ and ReS$_2$, are used for generating the lattice strain maps (Figure S4). Mapped from Figure 3a, the distributions of lattice strains along horizontal ($\varepsilon_{xx}$) and vertical ($\varepsilon_{yy}$) directions as well as shear strain ($\varepsilon_{xy}$) and rotation components are shown in Figure 3c-f, respectively. Note that the unstrained MoS$_2$ matrix has been set to be the zero-strain reference for both MoS$_2$ and ReS$_2$ regions. The precisions of the lattice strain maps are 0.005, 0.010, 0.006 and 0.006 for $\varepsilon_{xx}$, $\varepsilon_{yy}$, $\varepsilon_{xy}$ and rotation, respectively. The dashed lines noted in the maps indicate the contours of the nanodot #2 in Figure 3b. In contrast to the apparent lattice spacing changes for the large ReS$_2$ nanodots (nanodots #1~3 with lateral size > 10 nm) (bright red) in the $\varepsilon_{xx}$ map (Figure 3c), those smaller ones (less than 5~6 nm) show almost negligible changes in lattice spacings compared with the reference MoS$_2$ matrix (almost white). Such a difference unveils the fact that smaller ReS$_2$ nanodots are more readily to accommodate to the surrounding MoS$_2$ lattices. The same trend can also be observed in the $\varepsilon_{yy}$ map (Figure 3d), showing the same phenomenon in the vertical direction. Figure 3e indicates that the shear component $\varepsilon_{xy}$ is not prominent for ReS$_2$ nanodots of all sizes, while Figure 3f shows the discernable rotation of lattice of ReS$_2$ for the nanodot #1 of the largest size. Besides, from Figure 3c and d, the blue contours (*i.e.* reduced lattice spacing values) around ReS$_2$ nanodots #2 reveal the compression of the adjacent MoS$_2$ lattices which have coherent interfaces with ReS$_2$, because ReS$_2$ has larger lattice parameters in all directions compared with MoS$_2$.

Different from the built-in strains reported in the WSe$_2$-MoS$_2$ lateral heterostructures,[24] the laterally confined ReS$_2$ nanodots exhibit multi-domain structures when nanodot sizes are larger than 6 nm (Figure 3b). The yellow dashed lines in Figure 3b mark the domain boundaries in nanodot #2 as an example. The inhomogeneous red color inside the nanodot #2 in Figure 3c and d reveals the variation of lattice spacing due to the change of crystallographic orientations of individual domains. Due to the different degrees of lattice mismatch in three crystallographic directions mentioned above which are generally parallel to the ReS$_2$-MoS$_2$ interfaces according to our observations, the multi-domain structure can effectively avoid large lattice strains from strain accumulation along one direction in a single large domain. The reduction of total strain energy in such way acts as the driving force for the multi-domain formation. Therefore, there is a critical size for the formation of single-domain ReS$_2$ nanodots. When the ReS$_2$ nanodots are larger than the critical size, the total strain energy overwhelms the domain wall energy, giving rise to the multi-domain structure. The typical size of ReS$_2$ nanodots for such strain-induced multi-domain formation is 5~6 nm according to our STEM observations. Based on this experimental critical size for multi-domain formation, the domain boundary energy has been estimated to be about 30 meV/Å (See Supporting Information for details), which is in good agreement with the calculated domain boundary energies for most of the distorted 1T phase TMDs, *e.g.,* 27, 46, 40, 22, 51 and 52 meV/Å for $MoS_2$, $MoSe_2$, $MoTe_2$, $WS_2$, $WSe_2$ and $WTe_2$, respectively.[25] Despite the fact that heavy-dose electron beam irradiation at elevated temperature is reported to excite domain reconstruction,[19] herein mismatch strains are considered to be the primary origin for the multi-domain formation in nanodots since it elevates the instability of the ground state configuration. Furthermore, not only multi-domain structures, mismatch dislocations can also form at $MoS_2$/$ReS_2$ interfaces when the matrix $MoS_2$ lattice cannot tolerant large mismatch strains of large nanodots, for example nanodot #1 and #3. The concentrated strains in the vicinity of dislocations are visible in the strain maps as indicated by the black arrow heads in Figure 3c. The structures of these mismatch dislocations are displayed in the zoom-in STEM images (Figure S5).

The inlay of 1L $ReS_2$ nanodots with a round shape in 1L $MoS_2$ matrix in a coherent lattice manner resembles the 2D version of the "cylinder-fitting-a-hole" problem, where the "thick-wall cylinder" model[26] can be readily utilized to describe the strain and stress distributions inside and outside a nanodot. The ideal 2D strain ($\varepsilon_{2D,ideal}$) in the $ReS_2$ nanodot based on such simplified model can be given by:

$$\varepsilon_{2D,\text{ideal}} = \frac{-2(1-\nu_R)}{E_R} \frac{a_R - a_M}{a_M} \left( \frac{1+\nu_M}{E_M} + \frac{1-\nu_R}{E_R} \right)^{-1}$$

(1)

where $E$ and $\nu$ are Young's modulus and Poisson ratio, and subscripts M and R represent 1L MoS$_2$ and 1L ReS$_2$, respectively. $a$ is the lattice parameter of MoS$_2$ or the effective lattice parameter of ReS$_2$, assuming that it is isotropic. (See Supporting Information for derivation.) Taking the values $E_M = 270$ GPa, $E_R = 222$ GPa, $\nu_M = 0.27$ and $\nu_R = -0.1$ for 1L MoS$_2$ and ReS$_2$,[27,28] the $\varepsilon_{2D,\text{ideal}}$ is evaluated to be $-0.029$, which is ideally independent on nanodot size. Figure 3g shows the plot of this $\varepsilon_{2D,\text{ideal}}$ (red line) together with the experimental 2D strain ($\varepsilon_{2D} = \varepsilon_{xx} + \varepsilon_{yy}$) in the ReS$_2$ nanodots against nanodot size. Note that the experimental $\varepsilon_{2D}$ here are the strains of the ReS$_2$ nanodots, which are calculated by using the pristine ReS$_2$ lattice as the reference and converted from the lattice strain mapping with MoS$_2$ lattice as reference in Figure 3c-f. Herein, each data point is obtained from the statistics on a ReS$_2$ nanodot with a specific size, *i.e.*, each 2D strain value is averaged over the whole area of the corresponding nanodot, and the error bar shows the deviation of the strain values inside the corresponding nanodot. Interestingly, different from the constant $\varepsilon_{2D,\text{ideal}}$ predicted by the continuum mechanical modeling, an obvious size dependence of $\varepsilon_{2D}$ can be observed when the nanodots are smaller than ~4 nm, showing the deviation from the theoretical model. As a result, larger dispersion of $\varepsilon_{2D}$ values for nanodots smaller than 2 nm is discernible (Figure 3g). Nevertheless, the trend of enhanced contraction of nanodots smaller than ~4 nm, compared with $\varepsilon_{2D,ideal}$, is clearly observable. Such an evident size-effect can hardly be explained by the continuum mechanical model with size-independent elastic moduli. In fact, it indicates possible changes of electronic structures when the nanodots are smaller than 3~4 nm, corresponding to the partial or full phase transition from distorted 1T to 2H as observed in Figure 2. In addition, the experimental $\varepsilon_{2D}$ for the nanodots with size of 4~10 nm is in good agreement with $\varepsilon_{2D,ideal}$, regardless of having either a single or multiple domains. Such observations indicate that the multi-domain formation only plays a minor role in the reduction of total 2D strains by homogenizing the distributions of anisotropic strains.

Raman scattering and photoluminescence spectra of the 1L $ReS_2$ nanodots-inlaid 1L $MoS_2$ samples are shown in Figure 4. Here, the samples with 1.3, 3.0, and 5.1 at.% $ReS_2$ nanodot inlays are fabricated with plasma treatment for 10, 20 and 30 minutes before the 2$^{nd}$ CVD step for $ReS_2$ deposition. Exemplary STEM images of the three samples along with the corresponding statistics on nanodot sizes can be found in Figure S6. As plasma treatment time is extended, the average sizes and the dispersity of dot-sizes increase, which is primarily resulted from the increased population of large holes during plasma treatment. In addition to the characteristic E' and A'$_1$ symmetry Raman modes from MoS$_2$ matrix at ~384 and ~404 cm$^{-1}$ (Figure 4a),[29] weak signals in lower frequency ranges gradually emerge as ReS$_2$ nanodot population increases, corresponding to groups of A$_g^1$ and A$_g^{4\sim6}$ (150~160 cm$^{-1}$), A$_g^7$ and A$_g^8$ (210~240 cm$^{-1}$), and A$_g^{11\sim14}$ (300~330 cm$^{-1}$) like peaks from ReS$_2$ (Figure S7).[30,31] Raman features of the MoS$_2$ matrix are also changed with the ReS$_2$ nanodots inlaying. Figure 4b displays the evolution of MoS$_2$ E' and A'$_1$ modes with ReS$_2$ substitution. The clear trend of stiffening of A'$_1$ mode up to 3 at.% Re is observed while the shift of E' is trivial. Figure 4c is the plot of E'/A'$_1$ intensity ratio against ReS$_2$ substitution percentage, showing a substantial drop of E'/A'$_1$ intensity ratio with ReS$_2$ substitution. These features are in line with the alleviation of the negative charge doping propensity of MoS$_2$,[32] indicating the possible electron trapping effect of the ReS$_2$ nanodots. In the evolution of photoluminescence spectra under 514.5 nm laser excitation (Figure 4d), three peaks corresponding to A exciton, B exciton, and A$^-$ trion from 1L MoS$_2$ [33,34] appears in the spectra of pristine as well as ReS$_2$ nanodot-inlaid 1L MoS$_2$ samples, while characteristic emission from ReS$_2$ cannot be resolved under the intense MoS$_2$ emission background. In the plot of peak positions in Figure 4e, a mild blue shift tendency is observed for both A and B exciton peaks, possibly due to the band structure modulation as well as the compressive strain induced by larger ReS$_2$ lattices (Figure 2), while change of A$^-$ trion position is negligible. Moreover, A/A$^-$ intensity ratio is enhanced as ReS$_2$ nanodots are inlaid in Figure 4f (upper panel), indicating a more neutral charge state in 1L ReS$_2$ nanodots-inlaid MoS$_2$, which is in agreement with Raman scattering results. Therefore, the optical spectroscopy characterizations suggest the effective electron transfer from 1L MoS$_2$ matrix to the 1L ReS$_2$ quantum dots. But, obvious change of the total photoluminescence intensity (normalized by the E' Raman peak) is not observable (Figure 4f lower panel). In addition, the Raman and PL behaviors are in the opposite of the electron-doping by isolated Re-doping in 1L MoS$_2$, suggesting a fundamentally different role of the quantum dots from Re dopants (Figure S8).[35]

Both the Raman and PL results demonstrate an enhanced electron trapping behavior as ReS$_2$ substitution percentage increases as discussed above. While, a saturation can be observed after the ReS$_2$ substitution percentage reaches ~3 at.% and further increasing the loading amount by increasing the number of larger nanodots (Figure S6) makes only weak or moderate changes to the spectra. Therefore, the smaller nanodots are found to have more significant electron-trapping effect compared with larger ones in term of the number of per unit area. As elucidated by the atomic structure and lattice strain analyses, smaller ReS$_2$ nanodots are accompanied with larger lattice-confinement strains. At the same time, compressive strains are known to reduce the bandgap of 1L ReS$_2$, which is originally narrower than 1L MoS$_2$ (Figure S2c), or even cause metallization of ReS$_2$.[36,37] Moreover, 2H phase ReS$_2$ is also predicted to have a metallic nature.[38] Therefore, the smaller ReS$_2$ nanodots are expected to have deeper potential wells than the larger ones, although the nanosized ReS$_2$ is no longer characterized by the Bloch states in perfect ReS$_2$ crystals. On the other hand, from basic quantum mechanics, any 2D potential well supports at least one bound state no matter how shallow the well-depth is, whereas the trapping effect weakens exponentially with the reduction of well-depth.[39,40] Consequently, the deep levels or narrowed bandgaps of the significantly strained or even phase-transformed small ReS$_2$ nanodots (<2 nm) are expected to show more significant electron trapping effects than larger nanodots and play the central role in the charge-trapping-like behaviors in the ReS$_2$ quantum dots which are inlaid in 1L MoS$_2$. Such size- and strain-engineered 0D nanodots in 2D materials may shed lights on the quantum confinement at low temperatures under appropriate electrical tuning of the Fermi level to minimize the background of normal conduction electrons. Moreover, the materials design of inlaid quantum dots can be employed in other 2D TMD systems for obtaining 0D quantum confinement in 2D materials.

CONCLUSION

In summary, we demonstrated the successful inlay of 1L $ReS_2$ quantum dots in 1L $MoS_2$ using a two-step CVD growth method together with plasma treatment. The size-dependence of $ReS_2$ nanodot morphology, atomic structure and strain distribution are unveiled and expounded by atomic-scale HAADF-STEM. The growth of $ReS_2$ quantum dots in 1L $MoS_2$ matrix is found to follow a dispersion-limited manner, showing advantages in forming atomically sharp $ReS_2/MoS_2$ interfaces. Raman scattering and photoluminescence measurements reveal the electron transfer from 1L $MoS_2$ matrix to 1L $ReS_2$ quantum dots, indicating the possible electron-trapping effect of the $ReS_2$ quantum dots, which shows potential for realizing 0D quantum confinement in 2D materials at low temperatures.

**METHODS**

*Synthesis of pristine 1L $MoS_2$.* The pristine 1L $MoS_2$ samples were synthesized using a low-pressure chemical vapor deposition method. Sulfur (1.0 g, Wako Pure Chemical Industry, purity 99%), $MoO_3$ (10 mg, Sigma Aldrich, purity >99.5%), and micro slide glass substrate (Matsunami Glass, S1214) were placed in different CVD furnace zones in a sequence from upstream to downstream of the Ar gas flow. The growth chamber was pumped down to a base pressure of $5\times10^{-2}$ mbar before turning on Ar flow to 500 sccm. Then, sulfur, MoO$_3$ and substrate were heated to 120°C, 520°C and 750°C respectively and kept for 40 min for MoS$_2$ growth, followed by gentle cooling down to room temperature in 20 min.

*Nanovoids fabrication by plasma treatment.* Nanovoids were introduced in the CVD-grown MoS$_2$ by air plasma treatment using a JEOL JIC-410 plasma ion generator.[17] The samples on glass substrates were placed close to and in the same distance to the plasma electrode. The plasma voltage was kept constant at 267 V for three different periods (10, 20 and 30 min) to create nanovoids with different sizes and distributions.

*Epitaxy of ReS$_2$ nanodot by LPCVD.* The voided 1L MoS$_2$ samples on the glass substrates were quickly transferred to the CVD furnace after plasma treatment. A clean quartz tube without Mo contamination from former MoS$_2$ growth was used to avoid redeposition of MoS$_2$ in this step. CVD setup was the same as MoS$_2$ growth but MoO$_3$ source and bare substrate were replaced by Re$_2$O$_7$ (5.0 mg, Strem chemicals, purity 99.99%) and substrate with voided MoS$_2$. Sulfur was heated to 120°C 10 min before the heating of Re$_2$O$_7$ and substrate. Then, Re$_2$O$_7$ and the voided MoS$_2$ were heated to 300°C and 700°C respectively followed by 30 minutes keeping for ReS$_2$ nanodot inlay before cooling down.

*Microstructure characterization.* As-fabricated samples were quickly transferred from glass substrates to holey carbon coated copper grids for TEM characterization. Note that as-grown samples can be lifted from glass substrates and float on the surface of small water droplets, and only ultra-pure water was used throughout the transfer process, which enables the clean transfer of the samples. STEM images were captured using a JEOL JEM-2100F TEM equipped with double spherical aberration (Cs) correctors (CEOS) for imaging and probing lenses. Observations were performed at the acceleration voltage of 200 kV and the spatial resolution of the STEM is ~0.1 nm. The collecting angle for HAADF-STEM is between 64 and 171 mrad. Deconvolution of images was performed using the DeConvHAADF software (HREM Research Inc.).[41] Strain analysis based on the geometric phase analysis (GPA) method[42] was performed using the FRWRtools plugin for DigitalMicrograph (www.physics.hu-berlin.de/en/sem/software/software_frwrtools).

*Raman and photoluminescence measurements.* Raman and photoluminescence spectra were acquired from the samples transferred to SiO$_2$/Si substrates by a confocal Raman spectroscopic system (Renishaw InVia RM 1000) with 514.5 nm excitation laser at the power of 5 mW and 2 mW, respectively.

ASSOCIATED CONTENT

**Supporting Information**.

Supporting Information on experimental results and discussion. This material is available free of charge on the ACS Publications website at DOI:

Additional HAADF-STEM images showing structures of the voids, incoherently inlaid ReS$_2$ nanodots, and lattice mismatch dislocations; descriptions on the GPA lattice strain mapping; distribution and statistics of the ReS$_2$ nanodots; zoom-in Raman spectra showing the ReS$_2$-like signals; comparison with the Raman and PL of Re-doped MoS$_2$; estimation of domain boundary energy; derivation of the ideal 2D strain (PDF)

AUTHOR INFORMATION

**Corresponding Author**

*E-mail: mwchen@jhu.edu

**Author Contributions**

M. W. C. designed the project. Z. W. performed the sample fabrication, TEM and Raman/PL characterizations, and data analyses. R. L. contributed to the TEM. I. J. and H. K. contributed to the CVD. M. W. C. and Z. W. wrote the manuscript. All the authors discussed the results and commented on the manuscript.

ACKNOWLEDGMENT

This work is sponsored by Whiting School of Engineering, Johns Hopkins University, USA and World Premier International (WPI) Research Center Initiative for Atoms, Molecules and Materials, MEXT, Japan.

**TOC graph**

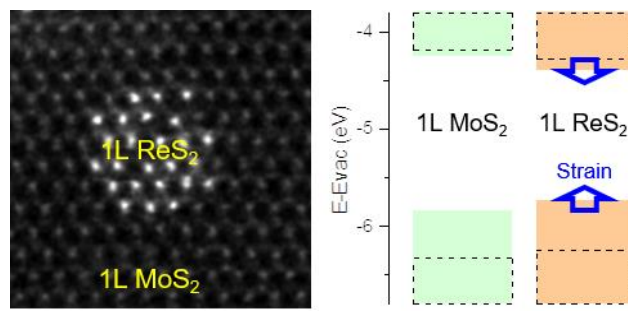

**Figures and captions**

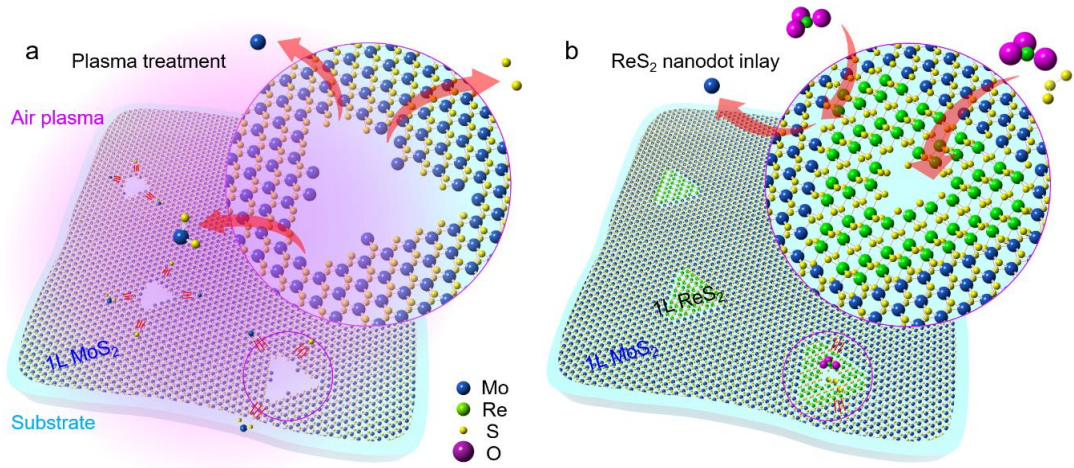

**Figure 1** Illustration on the nanodot fabrication techniques. (a) Plasma treatment for voids and defects creation. 1L MoS$_2$ samples grown on glass substrates by the first step CVD are exposed to air plasma. Under the bombardment of species in the air plasma, Mo and S atoms escape from the MoS$_2$ lattice, giving rise to voids randomly distributed in the 1L MoS$_2$ matrix. (b) ReS$_2$ nanodots inlay by the second step CVD where only Re-containing precursor and S are used. Lateral epitaxy of ReS$_2$ along the inner edges fills the voids and results in the final 1L ReS$_2$ nanodot in 1L MoS$_2$ matrix structure.

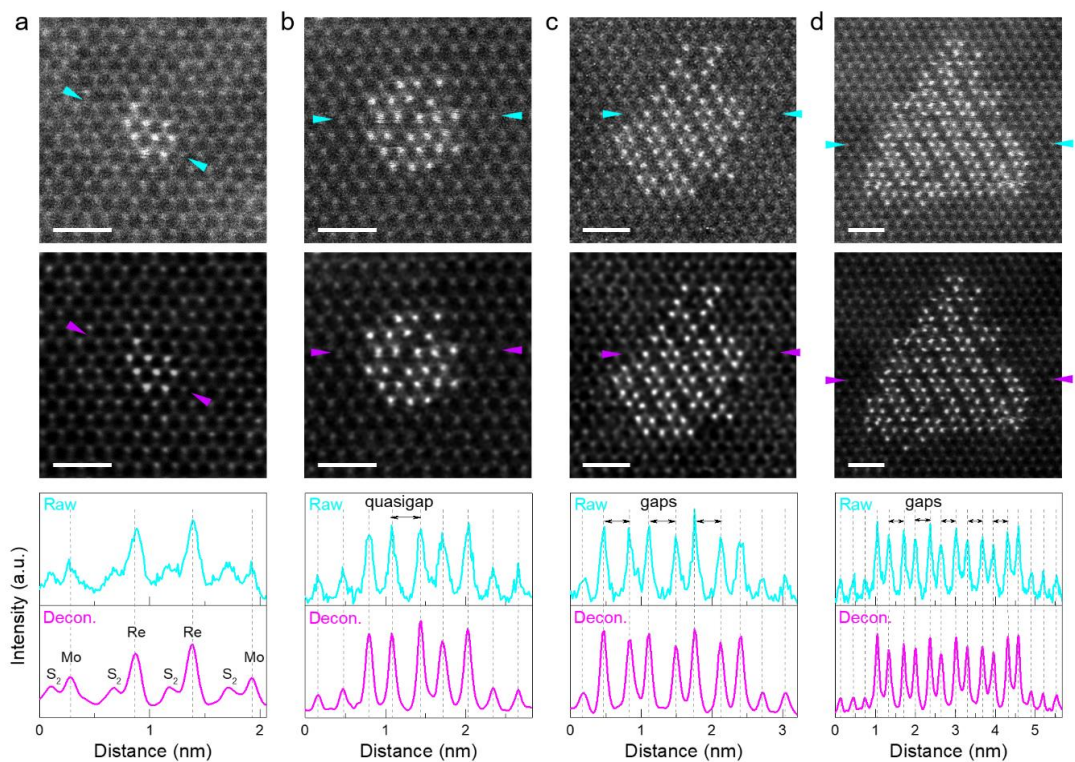

**Figure 2** Evolution of the atomic structure with size of $ReS_2$ nanodots. (a-d) Typical HAADF-STEM images of $ReS_2$ nanodots from subnano to ~5 nm to show the structure evolution with size. Upper panels are for raw images and middle ones for deconvoluted images of the same regions (scale bar: 1 nm). Lower panels are intensity profiles along the lines between two arrow heads in corresponding raw and deconvoluted images. Upper panels are for raw images and lower ones for deconvoluted images.

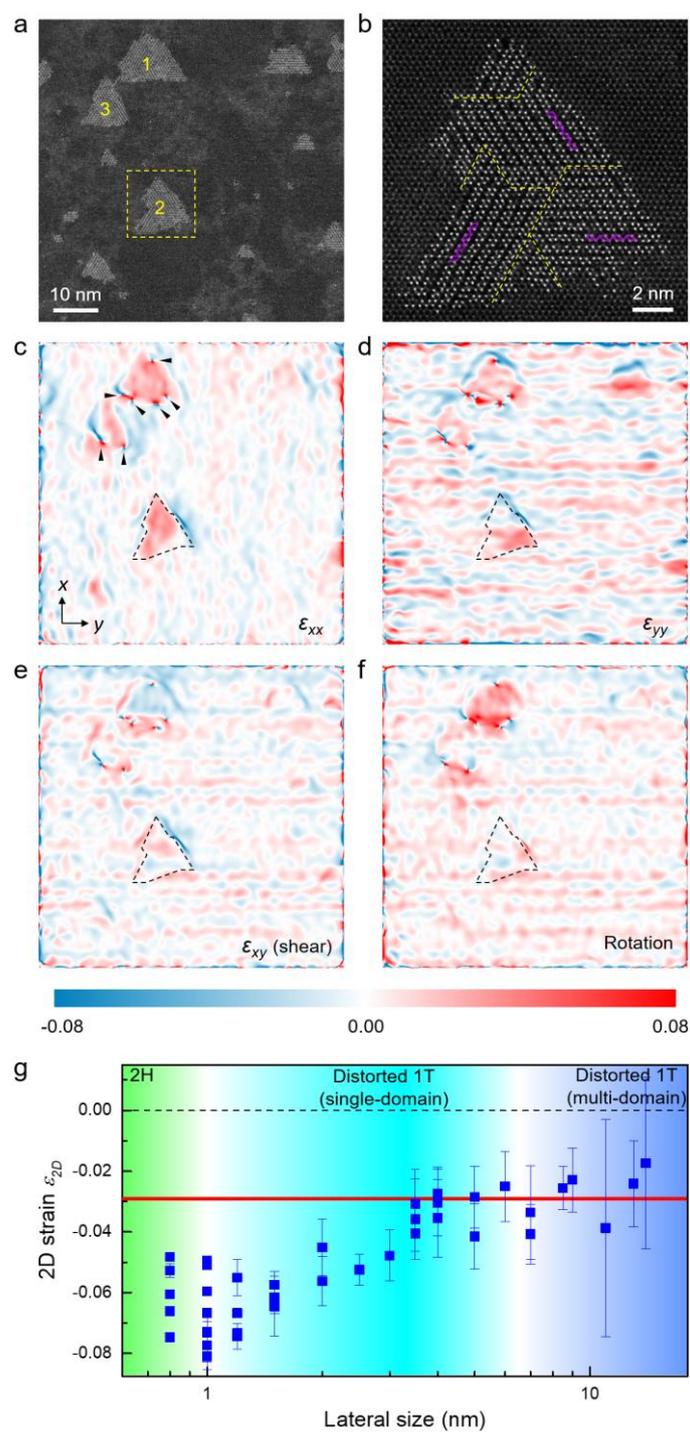

**Figure 3** Size-dependent straining behavior and domain structure in larger ReS$_2$ nanodots. (a) HAADF-STEM image of a region including larger and smaller ReS$_2$ nanodots. Three largest nanodots are numbered. (b) Zoom-in image of the region

indicated by the yellow dashed square in (a) revealing the multi-domain structure of nanodot #2. The yellow dashed lines indicate domain boundaries and the purple diamond chains show the domain directions. (c-f) $\varepsilon_{xx}$, $\varepsilon_{yy}$, $\varepsilon_{xy}$ and rotation strain mapping of (a) taking the plain 1L MoS$_2$ lattice as reference. Black arrow heads indicate dislocations in large nanodots in (c), which are omitted in (d-f). Black dashed lines in (c-f) shows the contour of the nanodot #2. Non-uniform blue and red colors inside the black dashed lines in (c,d) resolve the multiple domains in nanodot #2. (g) Size-dependence of 2D lattice strains in the ReS$_2$ nanodots with the pristine 1L ReS$_2$ as reference. Red line shows the ideal 2D strain derived from the macroscopic model.

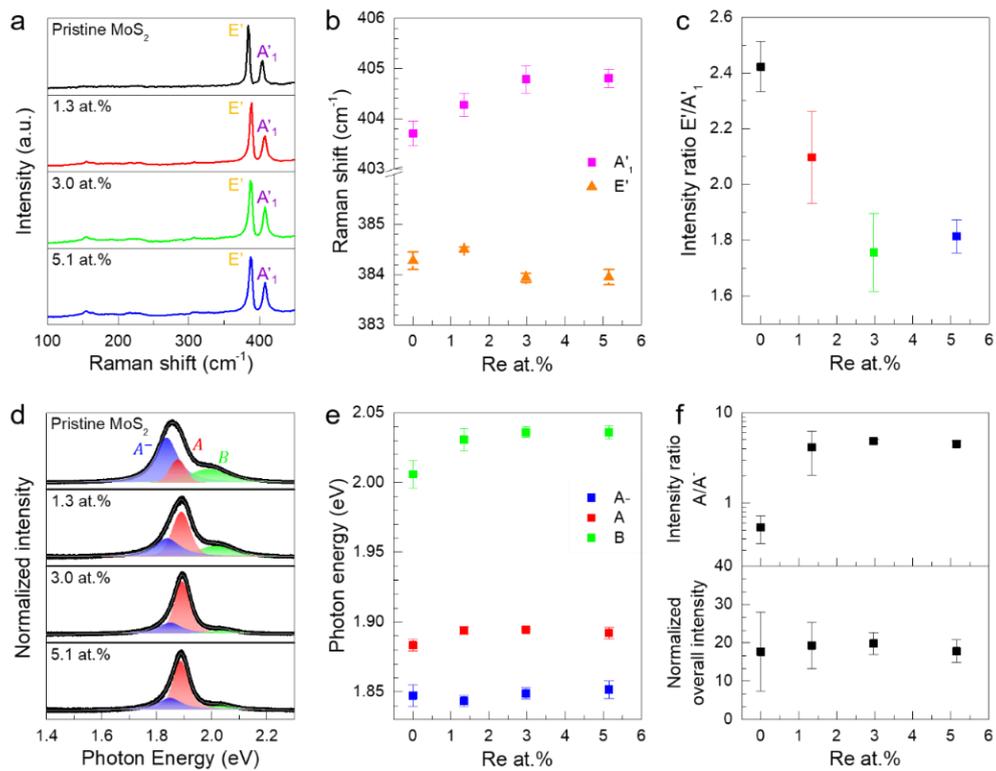

**Figure 4** Raman scattering and photoluminescence measurements on the 1L ReS$_2$ nanodot-inlaid MoS$_2$ samples. (a) From top to bottom: Raman spectra of pristine 1L MoS$_2$ and ReS$_2$ nanodot-inlaid MoS$_2$ samples of 1.3, 3.0, and 5.1 at.% ReS$_2$ nanodot substitution, corresponding to the samples fabricated by plasma treatment for 10, 20 and 30 minutes, respectively. Evolution of (b) peak positions of the MoS$_2$ E' and A'$_1$ modes and (c) the E'/A'$_1$ intensity ratio with ReS$_2$ substitution percentage. (d) From top to bottom: photoluminescence spectra of the same pristine 1L MoS$_2$ and ReS$_2$ nanodot-inlaid MoS$_2$ samples as in Raman scattering measurements. Evolution of (e) peak positions of A$^-$ trion and A and B excitons, upper (f) A/A$^-$ intensity ratio, and lower

(f) overall intensity normalized by the E' Raman peak as ReS$_2$ substitution percentage increases.